\input{aipcheck}

\documentclass[
,draft            
]{aipproc}
\usepackage{epsfig}
\usepackage{graphicx}
\usepackage{bm}
\layoutstyle{6x9}

\begin{document}
\title{Symmetry breaking in non-commutative cut-off field theories}


\author{Paolo Castorina and Dario Zappal\`a}{
  address={Dept. of Physics, University of Catania and INFN, Sezione di Catania- Italy}}

\begin{abstract}
The relation between symmetry breaking in non-commutative cut-off field theories and transitions 
to inhomogeneous phases in condensed matter and in finite density QCD is discussed.
The non-commutative dynamics, with its peculiar infrared-ultraviolet mixing, can be regarded as an effective description of the mechanisms which 
lead to inhomogeneous phase transitions and a  roton-like excitation spectrum.
\end{abstract}

\maketitle

\newcommand \be {\begin{equation}}
\newcommand \bea {\begin{eqnarray}}
\newcommand \ee {\end{equation}}
\newcommand \eea {\end{eqnarray}}

The relation between symmetry breaking in quantum field theory and phase transitions in condensed matter is well known.
The previous correspondence is usually restricted to the case of constant
order parameter  because this
guarantees the translational and rotational invariance of the field theory.
On the other hand, in condensed matter and, more recently, in high density QCD  one considers transitions 
from homogeneous to inhomogeneous phases with non-constant order parameters.
For a bosonic system, according to \cite{braz}, this kind of  transition is associated with an order parameter
which for large distances is an oscillating function and with  a roton-like behavior of the excitation spectrum.
For the fermionic systems it is possible to build inhomogeneous superconducting states,  
with energy  lower than the BCS state, where the particle-particle (p-p) Cooper pairs or the particle-hole (p-h) pairs
have  non-zero total momentum and then the corresponding fermionic condensates
are not uniform.
The former phase is described  in condensed matter systems by  the LOFF state \cite{loff} 
and in QCD is related with a diquark condensation with interesting theoretical and phenomenological implications \cite{raja,beppe}.
The latter has been originally proposed in condensed matter by  Overhauser \cite{over} (the (p-h) instability is called spin density wave)
and in QCD the corresponding phase, called chiral density wave, can compete with the QCD-BCS phase 
in the strong coupling regime and in an intermediate region of the chemical potential \cite{cdw}.

In \cite{gubser,noi1,noi2,noi3,bieten} it has been shown that the previous condensed matter 
and QCD inhomogeneous transitions  are also  typical features of 
quantum field theory with  non-commutative coordinates, $[x_\mu,x_\nu] = i \theta_{\mu\nu}$.
 As discussed in this talk,
 in the non-commutative generalization of $\lambda \phi^4$ theory
\cite{gubser,noi1,bieten}  the spontaneous symmetry 
breaking occurs for a non-uniform stripe phase and in the non-commutative Gross-Neveu (GN) model there is an
inhomogeneous chiral symmetry breaking corresponding to spin density waves \cite{noi2}.
These results are mainly  due to the infrared/ultraviolet (IR/UV) connection which characterizes the non commutative field theories
(for a review see \cite{iruv}).

We shall first discuss the relation between the symmetry breaking in  non-commutative 
self-interacting scalar field theory and the roton excitation in BEC and then we consider the fermionic systems.
On general grounds, in a bosonic condensate the roton spectrum is due to a non-local interatomic  potential 
$V(\vec r - \vec r\,')$, with a momentum dependent Fourier
transform. Since the BEC with the local (pseudo)potential $\delta(\vec r- \vec r\,')$ is analogous to the spontaneous symmetry breaking
in $\lambda \phi^4$ theory, one can assume that some relevant physical effects due to the
non-local repulsive interaction can be described
by generalizing the self-interacting field theory in such a way to introduce an effective non-local coupling.

 A simple approach is to consider the non-commutative $\lambda \phi^4$ theory with action
\be
\label{uno}
S(\phi)=\int d^4 x \left ({1 \over 2}\partial_{\mu} \phi ~ \partial^{\mu} \phi
- {1 \over 2} m^2 \phi ^2 -{\lambda
\over 4!} \phi ^{4*}\right )
\ee
where the star (Moyal)product is defined by ($i,j =1,.,4$)
\bea
\phi ^{4*}(x)= \phi(x) * \phi(x) * \phi(x) * \phi(x) = \hspace{70 pt}\nonumber\\
\label{due}
\exp\left \lbrack{ i \over 2} \sum_{ij}\theta_{\mu \nu} \partial^{\mu}_{x_i} \partial^{\nu}_{x_j}\right \rbrack
\Bigl (\left. \phi (x_1) \phi (x_2) \phi (x_3) \phi (x_4)
\Bigr )
\right |_{x_{i}=x}
\eea
The ``deformation'' of the self-interaction term by the Moyal product gives a 
momentum dependent repulsive effect which is responsible,
as we shall see below, for the roton spectrum and for the phase transition to an inhomogeneous background.
In \cite{noi1}, for  $\theta_{ij}= \epsilon_{ijk}\theta^k$ with $\vec \theta = (0,0,\theta)$, the spontaneous symmetry breaking 
for the theory in Eqs. (\ref{uno}) and (\ref{due}) has been analyzed with the following results:
{\bf 1)} the transition occurs to a stripe phase where the order parameter is $\phi(\vec x) =A \cos \vec Q \cdot \vec x$;
{\bf 2)} $A,Q$ and the energy excitation $\omega(p)$ are fixed by minimizing the energy;
{\bf 3)} $\vec Q$ is orthogonal to $ \vec \theta$ and $Q$ is small for large $\theta$;
{\bf 4)} the excitation spectrum can be approximated by
$\omega^2(\vec p)= p^2 + M^2(\vec p)$
where the function $M(\vec p)$ will be discussed later.
As discussed in detail in \cite{noi1}, since $Q$ is small, the inhomogeneous background is a smooth function of $x$ and then,
the breaking of translational (smooth) and rotational invariance is approximated by a translational invariant propagator with a
momentum dependent mass term.
In the particular case  $\vec Q = (Q/\sqrt{2}, Q/\sqrt{2},0)$ and large values of $\theta\Lambda^2$ ($\Lambda$ is the UV cut-off),
it turns out that $Q^2/ \Lambda^2 = (\lambda/24 \pi^2 )^{1/2} (1/ \theta \Lambda^2)$
and that $M(\vec p)$, for small $p$, is given by
\be
\label{cinque}
M^2(\vec p)\left. \right|_{p\to 0} \sim \alpha 
+{ {\lambda} \over {6\pi^2 }}\frac{1}{|\vec p \times \vec \theta|^2}
\ee
where $\alpha$ is a constant and $\times$ indicates the usual vector product.
The peculiar behavior for small $p$ of the last term in the previous equation is due to the IR/UV connection
of the non-commutative field theory and gives a divergent mass term
in the IR region and a minimum in the irreducible two-point function. 
However the effective theory has a natural self-generated IR cut-off $Q$ where it is more correct to cut
the small momenta and then  the excitation spectrum  should be correctly identified 
 for $p \geq Q$ by the previous expressions for $\omega(p)$ and $M^2(p)$. It has a roton-like dip 
at a typical scale of order $Q$.

The Moyal-deformed term in Eq. (\ref{uno}) can mimic 
effective interactions which are non-local and globally repulsive.
Then,  
the previous results of the non-commutative theory can describe some 
interesting physical effects of the Bose -Einstein trapped condensates
where the non-contact repulsive interaction is dominant.
This analysis has been performed in detail in ref. \cite{noi3} where 
the correlation between roton-like spectrum and non-uniform background of the non commutative theory has been 
compared with
the results obtained for BE trapped condensates, in analogous dynamical conditions. 

Let us now consider the informations coming from non-commutative effective field theories for 
fermionic systems.
 In \cite{noi2} the transition from homogeneous to inhomogeneous phase
has been obtained by generalizing the GN model, in four dimensions, to the non-commutative case with lagrangian
\be
\label{nove}
L(x)= i \bar \psi_\alpha \partial \hskip -0.2 cm
\slash \psi_\alpha + g \bar \psi_\alpha * \psi_\alpha *\bar \psi_\beta* \psi_\beta
- g \bar \psi_\alpha * \bar \psi_\beta * \psi_\alpha* \psi_\beta\; .
\ee
For $g$ larger than some critical value, one finds, as in tha commutative case, chiral symmetry breaking but, this time, in an inhomogeneous phase
where the pair correlation function has a dependence on a total momentum, $\vec P$ of the ("Cooper") pair,
with $P/ \Lambda \simeq (1/\theta \Lambda ^2)$.The order parameter turns out to be an oscillating function of
$\vec x$ and one has the breaking of translational, rotational and chiral
invariance : $<\bar \psi(x) \psi(x)> = [ 1+c\, P^2 {\rm cos}  ( Px )] <\bar \psi \psi>_0$, 
where $ <\bar \psi \psi>_0$ is the constant order parameter of the commutative case and $c$ is a numerical constant.
Also in this case \cite{noi2} the spectrum has roton-like dip in the plane orthogonal to $\vec \theta$ and one recovers 
the dynamical relation with the non-uniform ground state \cite{braz}.
The previous field theoretical model is then analogous to a system with a non-local, strong 
four-fermion interaction, with an inhomogeneous phase where the particle-hole (p-h) pairs have non-zero total 
momentum.As discussed, this  phase in QCD has an energy close to the BCS phase only for strong coupling ,
 corresponding to intermediate value of the density, that is before entering  the perturbative regime 
where the QCD- LOFF phase is realized. 

Finally, the non commutative field theoretical results indicate that the phase transitions to inhomogeneous
condensates are first order and one expects similar behavior for the corresponding condensed matter and QCD  systems.


\begin{thebibliography}{99}

\bibitem{braz} S.A. Brazovskii, Zh. Eksp. Teor. Fiz. {\bf 68} (1975)175.
\bibitem{loff} A. J. Larkin, Y. N. Ovchinnikov, Zh. Exsp. Theor. Fiz. 
{\bf 47} (1964) 1136; P. Fulde and R.A. Ferrell, Phys. Rev. {\bf 135} (1964) A550. 
\bibitem{raja} K. Rajagopal and F. Wilczek, ``Handbook of QCD'', Edited by M. Shifman, World Scientific 2001.
\bibitem{beppe} R. Casalbuoni and G. Nardulli,
Rev. Mod. Phys. {\bf 76} (2004) 263.
\bibitem{over}
A.W.Overhauser  Phys. Rev. {\bf 128} (1962) 1437.
\bibitem{cdw} 
R. Rapp, E. Shuryak and I. Zahed, Phys. Rev. {\bf D63} (2001) 034008.
\bibitem{gubser}
S.S. Gubser and S.L. Sondhi, Nucl. Phys. {\bf B605} ;(2001) 395.
\bibitem{noi1}
P.Castorina, and D. Zappal\`a, Phys. Rev. {\bf D68} (2003) 065008.
\bibitem{noi2}
P. Castorina, G. Riccobene and D. Zappal\`a,  Phys. Rev. {\bf D69} (2004) 105024.
\bibitem{noi3}
P. Castorina, G. Riccobene and D. Zappal\`a, hep-th
\bibitem{bieten}
J. Ambjorn, S. Catterall, Phys. Lett. {\bf B549} (2002) 253; W. Bietenholz, F. Hofheinz, J. Nishimura, Nucl. Phys. Proc. Suppl. 119 (2003) 941; 
Fortsch. Phys. 51 (2003) 745; e-Print Archive: hep-th/0404020.
\bibitem{iruv}
M. R. Douglas and N. A. Nekrasov, Rev. Mod. Phys. {\bf 73} (2001) 977;
R.J. Szabo, Phys. Rep. {\bf 378} (2003) 207.



\end{thebibliography}
\end{document}